\documentclass{article}

\usepackage{amsmath}
\usepackage{amsfonts}
\usepackage{graphicx}
\usepackage{mathrsfs}
\usepackage{algorithm2e}
\usepackage{fullpage}
\usepackage{authblk}
\usepackage[colorlinks,allcolors=blue]{hyperref}

\usepackage{color}

\renewcommand{\phi}{\varphi}
\renewcommand{\epsilon}{\varepsilon}
\newcommand{\by}{\mathbf{y}}

\newcommand{\OO}{\mathcal{O}}
\newcommand{\RR}{\mathbb{R}}

\newcommand{\PP}{\ensuremath{\textrm{P}}}

\newcommand{\CP}[2]{\PP\left(\left. #1 \, \right| #2 \right)}

\newcommand{\EE}{\ensuremath{\operatorname{E}}}

\newcommand{\bra}[1]{\langle #1 \rangle}         
\newcommand{\BK}[1]{ {\left( #1 \right)} }       
\newcommand{\sqBK}[1]{ {\left[ #1 \right]} }      
\newcommand{\curBK}[1]{ {\left\{ #1 \right\}} }

\newcommand{\norm}[1]{\left\Vert #1 \right\Vert}

\renewcommand{\hat}{\widehat}

\newcommand{\diag}{\mathsf{diag}}

\newcommand{\Normal}[1]{\ensuremath{\operatorname{N}}\BK{#1}}
\newcommand{\Poisson}[1]{\ensuremath{\operatorname{Pois}}\BK{#1}}

\newcommand{\Bernoulli}[1]{\operatorname{Bernoulli}\BK{#1}}

\newcommand{\dist}{\mathsf{dist}}
\newcommand{\dTV}{\mathsf{d}_{\mathsf{TV}}}

\newcommand{\Uniform}[1]{\ensuremath{\operatorname{Uniform}}\BK{#1}}

\newcommand{\resampling}{\ensuremath{\operatorname{Resampling}}}
\newcommand{\probasimplex}{\mathscr{P}}
\newcommand{\coupling}{\mathscr{C}}
\newcommand{\cost}{\mathsf{C}}

\begin{document}

\title{On Coupling Particle Filter Trajectories}

\author{Deborshee Sen}
\author{Alexandre H. Thiery}
\author{Ajay Jasra}
\affil{Department of Statistics \& Applied Probability \\ National University of Singapore}

\date{}

\maketitle

\begin{abstract}
Particle filters are a powerful and flexible tool for performing inference on state-space models. They involve a collection of samples evolving over time through a combination of sampling and re-sampling steps. The re-sampling step is necessary to ensure that weight degeneracy is avoided. In several situations of statistical interest, it is important to be able to compare the estimates produced by two different particle filters; consequently, being able to efficiently couple two particle filter trajectories is often of paramount importance. In this text, we propose several ways to do so. In particular, we leverage ideas from the optimal transportation literature. In general, though, computing the optimal transport map is extremely computationally expensive; to deal with this, we introduce computationally tractable approximations to optimal transport couplings. We demonstrate that our resulting algorithms for coupling two particle filter trajectories often perform orders of magnitude more efficiently than more standard approaches.

\end{abstract}

%
%

\section{Introduction} \label{sec:introduction}
Hidden Markov models (e.g. \cite{cappe2009inference}) are of widespread importance throughout science and engineering and have been studied under different names including state-space models (e.g. \cite{durbin2012time}) and dynamic models (e.g. \cite{harrison1999bayesian}). Some applications include ecology (e.g. \cite{newman2009monte}), epidemiology (e.g. \cite{king2008inapparent}), fault detection \cite{yin2015intelligent}, finance (e.g. \cite{johannes2009optimal}), medical physics \cite{ingle2015ultrasonic}, multitarget tracking \cite{sutharsan2012optimization}, and reliability prediction \cite{wei2013dynamic}.

A hidden Markov model (HMM) with measurable state space $ (\mathsf {X},\mathcal{X} )$, parameter $\theta \in \Theta$ 
and observation space $ (\mathsf{Y},\mathcal{Y} )$ is a process $ \{  (X_{n},Y_{n} );n\geq0 \} $ where $ \{ X_{n};n\geq0 \} $ is a Markov chain on $\mathsf{X}$, and each observation $Y_{n}$, valued in $\mathsf{Y}$, is conditionally independent of the rest of the process given $X_{n}$. Let $\mu_{\theta}$ and $f_\theta$ be respectively a probability distribution and a Markov kernel on $(\mathsf{X},\mathcal{X} )$, and let $g_\theta$ be a Markov
kernel acting from $ (\mathsf{X},\mathcal{X} )$ to $(\mathsf{Y},\mathcal{Y} )$, with $g_\theta(x,\cdot)$ admitting a strictly positive density, denoted similarly by $g_\theta(x,y)$, with respect to some dominating $\sigma$-finite
measure. The hidden Markov model specified by $\mu_{\theta}$, $f_\theta$ and $g_\theta$ is 
\begin{equation} \label{eq:HMM}
\left\{
\begin{aligned}
X_{0} &\sim\mu_{\theta}(\cdot), \\
X_{n} &\vert \{X_{n-1}=x_{n-1}\}  \sim f_\theta(x_{n-1},\cdot), ~ n \geq1, \\
Y_{n} &\vert \{ X_{n}=x_{n} \} \sim g_\theta(x_{n},\cdot), ~ n\geq 0.
\end{aligned}
\right.
\end{equation}
In the remainder of this text, we fix a time horizon $T \geq 1$. For a sequence of observations $y_{0:T} \equiv \{y_0, \ldots, y_T\}$, write $\nu_{\theta,T}(x_T)$ for the conditional distribution of $X_{T} | y_{0:T}$ and $\ell(\theta)$ for the log-likelihood; the likelihood equals
\begin{align*} 
\int_{\mathsf{X}^{T+1}} &\mu_{\theta}(x_0) \, g_\theta(x_0,y_0) 
\prod_{t=1}^T f_\theta(x_{t-1},x_t) \, g_\theta(x_t, y_t) \, d x_{0:t} \, .
\end{align*}
The distribution $\nu_{\theta,T}(x_T)$ is known as the \emph{filtering distribution}. 

Except in a few situations (e.g. finite state spaces or Gaussian models), the log-likelihood as well as the filtering distribution are intractable and have to be estimated numerically. The particle filter algorithm, known as the ``bootstrap'' algorithm \cite{gordon1993novel}, is a Monte-Carlo algorithm that involves a collection of particles evolving over time through a combination of sampling and re-sampling steps; we refer the reader to \cite{doucet2009tutorial} for a recent and very readable account on the subject. The particle filter can be used to produce an estimate $\hat{\ell}(\theta)$ of $\ell(\theta)$ and an estimate $\hat{\nu}_{\theta,T}(x_T)$ of $\nu_{\theta,T}(x_T)$. We sometimes write $\hat{\ell}^{\{N\}}(\theta)$ and $\hat{\nu}^{\{N\}}_{\theta,T}(x_T)$  to stress that the stochastic estimates to the log-likelihood $\ell(\theta)$ and the filtering distribution $\nu_{\theta,T}(x_T)$ have been obtained through a particle filter that employs $N \geq 1$ particles.

Note that in practice, the re-sampling step is done adaptively, only when a measure of particle diversity such as the effective sample size (ESS) \cite{kong1994sequential} falls below a predetermined threshold. For clarity of exposition, we only describe the algorithms presented in this text with a re-sampling step at each iteration; however, our numerical examples use an adaptive re-sampling strategy which we describe in the appropriate sections while presenting them.

%
%
\begin{algorithm}[h] 
\caption{Bootstrap Particle Filter.}
\label{alg:bootstrap}

\begin{itemize}

\item[] \textbf{For} $t=0$: 

\begin{itemize}

\item[] Sample $(X_{i,0} )_{i=1}^{N}$ independently from  $\mu_{\theta}$.

\item[] Report $\hat{Z}_{0} = (1/N) \sum_{i=1}^{N} g_\theta(X_{i,0}, y_{0})$.

\end{itemize}

\item[] \textbf{For} $1 \leq t \leq T$:

\begin{itemize}

\item[] Sample $ (X_{i,t} )_{i=1}^{N} \mid
(X_{i,t-1} )_{i=1}^{N}$ independently from 
$\frac{\sum_{i=1}^{N} g_\theta(X_{i,t-1}, y_{t-1}) \, f_\theta(X_{i,t-1},\cdot )}{\sum_{i=1}^{N} g_\theta(X_{i,t-1}, y_{t-1})}.$

\item[] Report $\hat{Z}_{t} = \hat{Z}_{t-1}  \times  \curBK{ (1/N) \sum_{i=1}^{N} g_\theta(X_{i,t}, y_{t})}$.
 
\end{itemize}

\item[] \textbf{Output}: $\hat{\ell}(\theta) \equiv \log \, \hat{Z}_T$ and 
$
\hat{\nu}_{T,\theta}(x_T) = \sum_{i=1}^{N} \frac{ g_\theta(X_{i,T}, y_{T}) } { \sum_{j=1}^{N} g_\theta(X_{j,T}, y_{T}) } \delta_{X_{i,T}} ( x_T ).
$%

\end{itemize}

\end{algorithm}

In the last line of Algorithm \ref{alg:bootstrap}, $\delta_{x}$ denotes the Dirac measure at $x \in \mathsf{X}$. There are many situations in which there is interest in comparing the value of the log-likelihood function $\ell(\theta)$ and the filtering distribution $\nu_{\theta,T}(x_T)$ at two different values of the parameters, $\theta \neq \theta'$; in other situations, the log-likelihood and filtering distribution of several probabilistic models for $\BK{X_t,Y_t}_{t \geq 0}$ need to be compared. In general, this typically involves concurrently running several particle filters. We now give several examples.
\begin{enumerate}
\item
{\bf Multi-level particle filters.} For estimating the log-likelihood $\ell(\theta)$, instead of running a single particle filter with a large number $N \geq 1$ of particles to obtain an approximation $\hat{\ell}^{\{N\}}(\theta)$, it is sometimes more efficient to use a telescoping decomposition of the type 
\begin{eqnarray*}
\ell_0(\theta) + \sum_{k=1}^K \left \{ \ell_{k}(\theta) - \ell_{k-1}(\theta) \right \} 
= \ell_0(\theta) + \sum_{k=1}^K \Delta \ell_k(\theta)
\end{eqnarray*}
where $\ell_k(\theta)$ designates the log-likelihood of the parameter $\theta \in \Theta$ associated to an approximate probabilistic model for $\{X_t,Y_t\}_{t \geq 0}$ and $\Delta \ell_k(\theta) := \ell_k(\theta) - \ell_{k-1}(\theta)$ is the delta log-likelihood between levels $k$ and $k-1$.
This idea \cite{giles2008multilevel} has appeared in the literature under the name of \emph{multi-level Monte Carlo} simulation; the reader is referred to \cite{giles2015multilevel} for a recent review of the state-of-the-art; see also \cite{jasra2016multilevel}. A recent stream of work has adapted these ideas to the context of particle filters \cite{hoel2015multilevel, beskos2015multilevel, del2016multilevel, jasra2015multilevel, jasra2016multilevel}. Thus, for estimating the filtering distribution $\nu_{\theta,T}(x_T)$, instead of running a single particle filter with a large number $N \geq 1$ of particles to obtain an approximation $\hat{\nu}^{\{N\}}_{\theta,T}(x_T)$, it is sometimes more efficient to use a telescoping decomposition of the type 
\begin{align*}
%
%
\nu_{0,\theta,T}(x_T) + \sum_{k=1}^K \left \{ \nu_{k,\theta,T}(x_T) - \nu_{k-1,\theta,T}(x_T) \right \},
\end{align*}
where $\nu_{k,\theta,T}(x_T)$ designates the filtering distribution associated to an approximate probabilistic model for $\{X_t,Y_t\}_{t \geq 0}$ . Typically, the approximate probabilistic models are of increasing accuracy as $k \to K$ but are also more computationally demanding to simulate. In the context of inference for diffusion processes, the approximate filtering distribution $\nu_{k,\theta,T}(x_T)$ may refer to an Euler-Maruyama discretization with time step $\Delta t \equiv  \delta / 2^k$, for some level-independent constant $\delta>0$; we refer the reader to \cite{jasra2015multilevel, jasra2016multilevel} for non-asymptotic analyses of multi-level particle filtering methods in this context. It is often the case that at low levels, that is, $k \ll K$, the filtering distribution $\nu_{k,\theta,T}(x_T)$ can be estimated extremely efficiently. The approximation $\hat{\nu}_{\textrm{ML};\theta,T}(x_{T})$ of the filtering distribution $\nu_{\theta,T}(x_{T})$ is given by
\begin{eqnarray*}
\hat{\nu}^{\{N_0\}}_{0,\theta,T}(x_T) 
+ \sum_{k=1}^K \left \{ \hat{\nu}^{\{N_k\}}_{k,\theta,T}(x_T) - \hat{\nu}^{\{N_k\}}_{k-1,\theta,T}(x_T) \right \}
\end{eqnarray*}
and the log-likelihood $\ell(\theta)$ is approximated by quantities of the type 
\begin{eqnarray*}
\hat{\ell}_{\textrm{ML}}(\theta) & \approx & \hat{\ell}_0^{\{N_0\}}(\theta) + \sum_{k=1}^K \left \{ \hat{\ell}_k^{\{N_k\}}(\theta) - \hat{\ell}_{k-1}^{\{N_k\}}(\theta) \right \}.
\end{eqnarray*}
The estimates $\hat{\ell}_k^{\{N_k\}}(\theta)$ and $\hat{\nu}^{\{N_k\}}_{k,\theta,T}(x_T)$ are obtained by running a standard particle filter with $N_k \geq 1$ particles. In many situations, the so called \emph{multi-level estimates} $\hat{\ell}_{\textrm{ML}}(\theta)$ and $\hat{\nu}_{\textrm{ML};\theta,T}(x_{T}) $ can achieve an accuracy similar to the standard estimates $\hat{\ell}^{\{N\}}(\theta)$ and $\hat{\nu}^{\{N\}}_{\theta,T}(x_T)$ with a number of particles $N_k$ at level $k$, for $k \approx K$, orders of magnitude less than $N$; this can result in important computational savings.

\item 
{\bf Gradient estimate.}
Consider the case where $\Theta \subset \RR^d$.
In many situations, the transition densities of the latent process $\{X_t\}_{t \geq 0}$ are unavailable while it is still possible to simulate realizations of it. In these cases, algorithms for maximum likelihood estimation (or maximum a posteriori) typically rely on finite-difference approximations of the gradient of the log-likelihood. The property that the model for the latent process enters the algorithm only through the requirement that realizations of it can be simulated at any value of the parameter has been called \emph{plug-and-play} \cite{ionides2011iterated} since the simulation code can simply be plugged into the inference algorithm. The gradient can be approximated by finite differences 
\begin{equation*}
\bra{\nabla \ell(\theta), e_i}  \approx \frac {\hat{\ell}(\theta+\epsilon \, e_i) - \hat{\ell}(\theta - \epsilon \, e_i)} {2 \, \epsilon}
\end{equation*}
for an orthonormal basis $\BK{e_1, \ldots, e_d}$ of $\RR^d$ and discretization parameter $\epsilon > 0$.
It is worth pointing out that when the transition densities of the latent process are available, much more efficient methods are available; see \cite{poyiadjis2011particle, nemeth2014sequential, kantas2015particle}. 

\item
{\bf Markov chain Monte Carlo.}
Several sophisticated Markov chain Monte Carlo (MCMC) approaches have recently been proposed for Bayesian inference in state space models. If $\rho(\theta) \, d \theta$ denotes a prior density on the parameter $\theta \in \Theta$ and $\mathsf{q}(\theta, \theta') \, d \theta'$ a Markov proposal kernel, the Markov chain Monte Carlo approaches described in \cite{o2000analyses, andrieu2009pseudo, andrieu2010particle} consider an acceptance probability of a proposed move $\theta \mapsto \theta'$ of the type 
\begin{equation} \label{eq.mcmc_acceptance_ratio}
\min\curBK{1, \frac{\rho(\theta') \, \mathsf{q}(\theta', \theta)}{\rho(\theta) \, \mathsf{q}(\theta,\theta')} \, e^{\Delta \, \hat{\ell}(\theta, \theta')}},
\end{equation}
where $\Delta \, \hat{\ell}(\theta, \theta') \equiv \hat{\ell}(\theta') - \hat{\ell}(\theta)$ is an estimate of the delta log-likelihood $\ell(\theta') - \ell(\theta)$.
It is by now well understood that the performances of such Markov chain Monte Carlo algorithms is dictated by the variability of the estimate $\Delta \, \hat{\ell}(\theta, \theta')$ to the delta log-likelihood \cite{andrieu2014establishing, andrieu2015convergence, doucet2015efficient, sherlock2015efficiency}.
\end{enumerate}
In all the above-mentioned examples, approximating the difference between two values of the log-likelihood function could be carried out by running two \emph{independent} particle filters. Nevertheless, it is often of paramount importance for computational efficiency to reduce as much as possible the variability of the estimate to the delta log-likelihood; this effectively means being able to efficiently couple the trajectories of two particle filters. We present in this text several strategies that provide orders of magnitude improvements over the naive strategy that consists of running two independent particle filters. To this end, we propose approaches based on optimal transport and describe ways to efficiently implement them through sparse matrix computations. Extensive numerical simulations comparing the different approaches for coupling particles filters, taking computational time into account, are presented in the Section \ref{sec:numerical_investigations} of the paper; our proposed methodology is up to three orders of magnitude faster than its competitors.

The application of the optimal transportation methodology to particle filtering is not new. In the seminal paper \cite{reich2013nonparametric}, the authors replace the standard re-sampling step of sequential Monte Carlo methods by an optimal transport problem; see also \cite{reich2013guided, gregory2016multilevel}. By exploiting the optimal transport approach, the authors are able to obtain state-of-the-art result for data-assimilation in high-dimensional systems.
In this paper, we leverage optimal transportation methodologies for a very different purpose: the efficient coupling of two particle filter trajectories. 
In a work independent from ours, the authors of \cite{jacob2016coupling} also consider optimal transport for the coupling of particle filters; this is used to develop new smoothing methods. The focus of our paper is different; we design scalable methods that can be employed for a large number of particles through the use of sparse linear algebra approaches.

\paragraph{Notations} $\RR$ denotes the real line $(-\infty,\infty)$ and $\RR_+$ denotes its non-negative part $[0, \infty)$. For real numbers $a$ and $b$, $a \wedge b$ denotes their minimum. For a number of particles $N \geq 1$, the vector of zeros (respectively ones) of length $N$ is denoted by $\mathbf{0}_{N}$ (respectively $\mathbf{1}_{N}$); we set $[N] = \{1, \ldots, N\}$ and $y_{p:q} \equiv \BK{ y_{p}, \ldots, y_{q}}$ for $p \leq q$. The notation $\Normal{\mu,\Sigma}$ designates the Gaussian distribution with mean $\mu \in \RR^d$ and covariance $\Sigma \in \RR^{d ,d}$. For two vectors $u,v \in \RR^d$, the usual inner product is denoted by $\bra{u,v}$ and we set $\norm{u}^2 \equiv \bra{u,u}$. For two matrices $M,N \in \RR^{d,d}$, the Frobenius scalar product is $\bra{M,N} \equiv \textrm{tr}({M^T \, N}) = \sum_{i,j} M_{i,j} \, N_{i,j}$. For two distributions $\mu$ and $\nu$ on the measurable space $\mathsf{X}$, the product $\mu \otimes \nu$ designates the trivial coupling such that $\mu \otimes \nu(A \times B) = \mu(A) \, \nu(B)$ for all measurable subsets $A$ and $B$ of $\mathsf{X}$.

\paragraph{Organization of paper} The rest of this paper is organized as follows. We discuss coupling of particle filters in Section \ref{sec:coupling_particle_filters}. In particular, we formalize the set-up in Section \ref{sec:formal_setup} and present in Algorithm \ref{alg.coupled.PF} a generic coupled particle filter. We describe the idea of a coupled re-sampling step within this context and in Sections \ref{sec:independent_resampling} and \ref{sec.maximal.coupling} present some coupling schemes. However, we demonstrate that these schemes can be improved upon, and Section \ref{sec:OT_resampling} introduces the idea of using optimal transport \cite{kantorovitch1958translocation} to perform a coupled re-sampling step. Since in general computing the optimal transportation distance is computationally expensive, Section \ref{sec.IPF} considers approximate solutions to the optimal transportion problem that scale quadratically in the total number of particles, and then Section \ref{sec.knn} further reduces the cost down to sub-quadratic in the total number of particles. We demonstrate numerically the benefit of performing a coupled re-samplng step in Section \ref{sec:numerical_investigations} using our proposed sub-quadratic approximate optimal transport solution. Finally, Section \ref{sec:conclusion} concludes. 

%
%

\section{Coupling of particle filters} \label{sec:coupling_particle_filters}

\subsection{Formal set-up} \label{sec:formal_setup}

In the remainder of this paper, we will use some slightly more generic notations to describe particle filters; this will ease the presentation of the algorithms to follow. A particle filter on the state space $(\mathsf{X},\mathcal{X})$ with time horizon $T \geq 1$ can be described by an initial density $\mu$, a sequence of Markov kernels $\{m_{t}(x,dx)\}_{t=1}^T$ and weight functions $\{g_{t}(x)\}_{t=0}^T$, where $g_t: \mathsf{X} \to (0, \infty)$ is assumed to be strictly positive for simplicity; the associated particle filter gives a way, among other things, to approximate the marginal likelihood
\begin{equation} \label{eq.normalization.constant}
\int_{\mathsf{X}^{T+1}} \mu(x_0) \, g_0(x_0) \prod_{t=1}^T m_t(x_{t-1},x_t) \, g_t(x_t) \, dx_{0:T},
\end{equation}
which we assume is finite.

We consider two particle filters with associated initial densities $\mu^{(j)}$, Markov kernels $m_t^{(j)}$ and weight functions $g_t^{(j)}$ with $j \in \{1,2\}$. For simplicity, we assume in the remainder of this text that the state space $\mathsf{X}$ is endowed with a distance $\dist(\cdot, \cdot)$, although more general settings can easily be accommodated. The first step when attempting to couple two particle filter trajectories is to use the same ``noise" to drive the two particle systems. To this end, let us consider the usual algorithmic description of a Markov kernel $m^{(j)}_t(x, dx)$: there is a function $M^{(j)}_t: \mathsf{X} \times [0,1] \rightarrow \mathsf{X}$ that is such that, if $U \sim \Uniform{[0,1]}$ and $x_{t-1} \in \mathsf{X}$ is a fixed element of $\mathsf{X}$, the random variable $X_t \equiv M^{(j)}_t(x_{t-1}, U)$ is distributed as $m^{(j)}_t(x_{t-1}, dx)$. The main underlying assumption for being able to efficiently couple the two particle filters is that, if $X^{(1)}_{t-1}$ and $X^{(2)}_{t-1}$ are two random variables that are highly coupled (e.g. the mean distance $E[ \dist(X^{(1)}_{t-1}, X^{(2)}_{t-1})]$ is small, although other notions of \emph{closeness} can be used instead) then, if $U \sim \Uniform{[0,1]}$ is a uniform random variable independent from all other sources of randomness, the two random variables $X^{(1)}_t$ and $X^{(2)}_t$ defined as $X^{(1)}_t \equiv M^{(1)}_t(X^{(1)}_{t-1}, U)$ and $X^{(2)}_t \equiv M^{(2)}_t(X^{(2)}_{t-1}, U)$ are also highly coupled. 

In order to efficiently couple two particle filter trajectories, it is also crucial to be able to  carry out a \emph{coupled re-sampling} step; the main purpose of this paper is to investigate efficient strategies to do so. Consider two weighted $N$-particles systems $(X^{(1)}, W^{(1)}) \equiv \{(X^{(1)}_i,W^{(1)}_i)\}_{i=1}^N$ and $(X^{(2)}, W^{(2)}) \equiv \{(X^{(2)}_i,W^{(2)}_i)\}_{i=1}^N$ in $(\mathsf{X},\mathcal{X})$; we implicitly assumed that $W^{(1)},W^{(2)} \in \probasimplex_N$ for $j \in \{1,2\}$, where the probability simplex is defined as 
\begin{equation*}
\probasimplex_N \equiv \curBK{(w_1, \ldots, w_N) \in [0,1]^N \, : \, w_1 + \cdots + w_N = 1}.
\end{equation*}
For ease of notation, we identify a vector of $\probasimplex_N$ with its associated probability distribution on $[N]$. A re-sampling scheme is any function 
\begin{equation*}
\resampling:\mathsf{X}^N \times \probasimplex_N \times \mathsf{X}^N \times \probasimplex_N \times [0,1] \to [N]^N \times [N]^N
\end{equation*}
such that, if $U \sim \Uniform{[0,1]}$ is a random variable, the two (random) uniformly weighted particles systems \\ $\{(\bar{X}^{(1)}_i,1/N)\}_{i=1}^N$ and $\{(\bar{X}^{(2)}_i,1/N)\}_{i=1}^N$ defined by first sampling the \emph{ancestors} vectors $a^{(1)},a^{(2)} \in [N]^N$,
\begin{equation*}
\BK{a^{(1)}, a^{(2)}} \equiv \resampling\BK{X^{(1)}, W^{(1)},X^{(2)}, W^{(2)},U},
\end{equation*}
and setting $\bar{X}^{(j)}_{i} = X^{(j)}_{a^{(j)}_i}$ are such that the following identity holds
\begin{align} \label{eq.good.re-sampling}
\EE \sqBK{ \frac{1}{N} \sum_{i=1}^N \phi \left ( \bar{X}^{(j)}_i \right )}
=
\sum_{i=1}^N W^{(j)} \, \phi \left ( X^{(j)}_i \right )
\end{align}
for $j \in \{1,2\}$ for any function $\phi$ for which the expectation is finite. We will describe in the sequel several choices of re-sampling schemes. With these notations introduced, a generic way of coupling two particle filter trajectories is described in Algorithm \ref{alg.coupled.PF}. 

%
%

\begin{algorithm}[h]
\label{alg.coupled.PF}
\caption{Generic Coupled Particle Filter.}

\begin{itemize}

\item[] \textbf{For} $t=0$:

\begin{itemize}

\item[] Sample $ (X^{(j)}_{i,0} )_{i=1}^{N} ~\text{independently from}~ \mu^{(j)}$ for $j \in \{1,2\}$.\\
Define $W^{(j)}_0 \in \probasimplex_N$ by setting $W^{(j)}_{i,0} \, \propto \, g^{(j)}_0(X^{(j)}_{i,0})$ and report $\hat{Z}^{(j)}_{0} = (1/N) \sum_{i=1}^{N} g^{(j)}_0(X^{(j)}_{i,0})$.

\end{itemize}

\item[] \textbf{For} $1 \leq t \leq T$:

\begin{itemize} 

\item[] Sample $U_{t} \sim \Uniform{[0,1]}$ and set the ancestor vectors as
\begin{equation*}
(a^{(1)}_t, a^{(2)}_t) = \resampling\BK{X^{(1)}_{t-1}, W^{(1)}_{t-1},X^{(2)}_{t-1}, W^{(2)}_{t-1},U_t}.
\end{equation*}

\item[] Sample $N$ i.i.d uniform random variables $\{ U_{i,t} \}_{i=1}^N$ and propagate the particles
\begin{equation*}
X^{(j)}_{i,t} \; = \; M^{(j)}_t \left ( X^{(j)}_{a^{(j)}_{i,t},t-1}, U_{i,t} \right ) ~~~ \text{ for } j \in \{1, 2 \}.
\end{equation*} 

\item[] Define $W^{(j)}_{t} \in \probasimplex_N$ by setting $W^{(j)}_{i,t} \, \propto \, g^{(j)}_t(X^{(j)}_{i,t})$ and report $\hat{Z}^{(j)}_{t} = \hat{Z}^{(j)}_{t-1} \times \{ (1/N) \, \sum_{i=1}^{N} g^{(j)}_t ( X^{(j)}_{i,t} ) \}$ for $j \in \{1, 2 \}$

\end{itemize}

\item[] \textbf{Output}: $\log \, \hat{Z}^{(j)}_T$ and $\sum_{i=1}^{N} W_{i,T}^{(j)} \,  \delta_{X_{i,T}^{(j)}}$ for $j \in \{1, 2 \}$. 

\end{itemize}

\end{algorithm}

The ancestry lineage vector $b^{(j)}_{i,[t]}=(b^{(j)}_{i,0,[t]}, \ldots, b^{(j)}_{i,t,[t]}) \in [N]^{t+1}$ of particle $X^{(j)}_{i,t}$ is defined recursively as $b^{(j)}_{i,t,[t]} = i$ and $b^{(j)}_{i,k-1,[t]} = a^{(j)}_{b^{(j)}_{i,k,[t]}}$ for $k=1, \ldots,t$. At stage $1 \leq t \leq T$, two particles $X^{(1)}_{i,t}$ and $X^{(2)}_{i,t}$ are said to be coupled if the two ancestry vectors $b^{(1)}_{i,[t]}$ and  $b^{(2)}_{i,[t]}$ are identical. It is important to note that if $X^{(1)}_{i,t}$ and $X^{(2)}_{i,t}$ are coupled, the two particles have been driven by the same noise from time 0 all the way to time $t$. The number
\begin{equation*}
C_t = \textrm{Card}\left \{ i \in [N] \, : \, b^{(1)}_{i,[t]} = b^{(2)}_{i,[t]} \right \}
\end{equation*}
of particles that are coupled at time $t$ is a decreasing function of $t$ and $C_0=N$. A natural strategy to efficiently couple two particle filters is to try to maximize the number of particles that stay coupled, that is, make the function $t \mapsto C_t$ decrease as slowly as possible. We nevertheless demonstrate that, under natural continuity assumptions on the algorithmic representations $M^{(j)}_j$ of the Markov kernels $m^{(j)}_t$, there are other strategies that are much more efficient.

All the re-sampling schemes described in this text work by first creating a coupling matrix. Given $\pi^{(1)}$ and $\pi^{(2)}$ two discrete probability distributions on $[N]$ represented by the vectors $W^{(1)}, W^{(2)} \in \probasimplex_N$, a coupling matrix $\Pi \in \mathbb{R}^{N \times N}$ is any matrix that represents a probability distribution $\pi$ on $[N] \times [N]$ and that has $W^{(1)}$ and $W^{(2)}$ as marginal distributions; in other words, for any $1 \leq i \leq N$ we have
\begin{equation} \label{eq.coupling.constraint}
\sum_{\alpha=1}^N \Pi_{\alpha,i} = W^{(1)}_i ~~ \textrm{and} ~~ \sum_{\alpha=1}^N \Pi_{i,\alpha} = W^{(2)}_i.
\end{equation}
Given two weighted particles systems $(X^{(1)}, W^{(1)})$ and \\ $(X^{(2)}, W^{(2)})$ and a coupling matrix $\Pi$ of $W^{(1)}$ and $W^{(2)}$, it is straightforward to construct re-sampling schemes. We describe two such possibilities.

\begin{enumerate}
\item 
{\bf Multinomial re-sampling.} The matrix $\Pi$ represents a probability distribution $\pi$ on $[N] \times [N]$. If $\{ (a^{(1)}_k,a^{(2)}_k) \}_{k=1}^N$ are $N$ independent and identically distributed samples from $\pi$, since $\PP(a^{(j)}_k = i) = W^{(j)}_i$ for $1 \leq j \leq 2$, the resulting vectors $a^{(j)} = (a^{(j)}_1, \ldots, a^{(j)}_N)$ do satisfy property \eqref{eq.good.re-sampling}.
\item
{\bf Systematic re-sampling.} Let $U \sim \Uniform{[0,1]}$ be a draw from a uniform random variable on $[0,1]$ and consider an arbitrary ordering $\{(\alpha_k, \beta_k)\}_{k=1}^{N^2}$ of \\ $\{(1,1), (1,2), \ldots, (N,N)\}$. For $1 \leq k \leq N$, set
\begin{eqnarray*}
\Phi(k) & = & \inf \bigg \{ 1 \leq i \leq N^2 \; : \; \qquad \sum_{j=1}^i \Pi_{\alpha_j, \beta_j} \geq (U/N) + (k-1)/N \bigg \}
\end{eqnarray*}
and define $(a^{(1)}_k, a^{(2)}_k) = (\alpha_{\Phi(k)}, \beta_{\Phi(k)})$. The proof that the standard systematic re-sampling \cite{kitagawa1996monte} is correct is immediately adapted to this slightly generalized setting to show that the vectors $a^{(j)} = (a^{(j)}_1, \ldots, a^{(j)}_N)$ do satisfy property \eqref{eq.good.re-sampling}.
\end{enumerate}
%


%
%
%
\subsection{Independent re-sampling} \label{sec:independent_resampling}
Independent re-sampling is the most straightforward, and, of those we will consider, the least efficient way of carrying out a coupled re-sampling. It simply consists of choosing the trivial coupling matrix 
$\Pi_{\alpha, \beta} = W^{(1)}_{\alpha} \, W^{(2)}_{\beta}$.
%
%
Indeed, the matrix $\Pi$ describes the probability distribution $W^{(1)} \otimes W^{(2)}$. Suppose that at time $t-1$ of Algorithm \ref{alg.coupled.PF} all the $N$ particles are coupled. After an independent re-sampling step, the expected number of paired particles at time $t$ equals $\sum_{i=1}^N \PP(a^{(1)}_{i,t} = a^{(2)}_{i,t}) = N \times \sum_{j=1}^N W^{(1)}_{j,t} \, W^{(2)}_{j,t}$. In the  (very favourable) case where $W^{(1)}_t$ and $W^{(2)}_t$ both represent the uniform distribution on $[N]$, the expected number of paired particles at time $t$ only equals one. This is an indication that the independent re-sampling scheme performs extremely badly. Even in the case where the two particle filters are identical (that is, same initial distribution, Markov kernels and weight functions), a situation where it should be straightforward to keep all the particles coupled at all time $0 \leq t \leq T$, using an independent re-sampling scheme leads to an algorithm where all the particles are decoupled after just a few re-sampling steps.

\subsection{Maximal coupling} \label{sec.maximal.coupling} 
Considering the undesirable behaviour of the independent re-sampling scheme, we now describe in this section another re-sampling scheme that aims at keeping the number of paired particles as high as possible. Constructing a coupling that maximizes the number of paired particles is equivalent to looking for a coupling $\pi$ between the probability distributions $\pi^{(1)}$ and $\pi^{(2)}$ such that if $(Y,Z) \sim \pi$, the quantity $\PP(Y=Z)$ is maximized; this is indeed equivalent to finding the coupling matrix $\Pi$ whose trace is maximal. There always exists such a coupling and, except in degenerate situations, this coupling is unique. Indeed, this coupling is nothing else than the standard so-called \emph{maximal coupling} (e.g. \cite{ross2007second}, Chapter 2) whose probabilistic description is as follows. For $W^{(1)}, W^{(2)} \in \probasimplex_N$, set
\begin{equation*}
p \equiv \sum_{\alpha=1}^N \min\BK{W^{(1)}_\alpha, W^{(2)}_\alpha} = 1 - \dTV\BK{ W^{(1)}, W^{(2)} } \in [0,1],
\end{equation*}
where $\dTV\BK{ W^{(1)}, W^{(2)} }$ designates the total-variation distance between $W^{(1)}$ and $W^{(2)}$.
Let $\mu,\mu^{(1)},\mu^{(2)} \in \probasimplex_N$ be given by $\mu(\alpha) = p^{-1} \, \min ( W^{(1)}_\alpha, W^{(2)}_\alpha )$ and, for $j \in \{1,2\}$,
\begin{equation*}
\mu^{(j)}(\alpha) = (1-p)^{-1} \, \BK{W^{(j)}_\alpha - \min\BK{W^{(1)}_\alpha, W^{(2)}_\alpha}}.
\end{equation*}
If $p=0$, the probability distribution $\mu$ is not defined but, as will be clear in a moment, that is not a problem. The maximal coupling between $W^{(1)}$ and $W^{(2)}$ is the law of the random variable
\begin{equation} \label{eq.max.coupl}
(Y,Z) = B \cdot (\Gamma, \Gamma) + (1-B) \cdot (\Gamma^{(1)}, \Gamma^{(2)}),
\end{equation}
where $\Gamma,\Gamma^{(1)},\Gamma^{(2)}$ are three independent random variables with respective laws $\mu, \mu^{(1)},\mu^{(2)}$ and $B \sim \Bernoulli{p}$ is independent from any other sources of randomness. In other words, with probability $p$ we have $Y=Z$ while with probability $(1-p)$ the random variables $Y$ and $Z$ are independent and, since $\mu^{(1)}$ and $\mu^{(2)}$ are singular, distinct. 

Contrary to the independent re-sampling scheme, if one uses the maximal coupling re-sampling for two identical particle filters, all the particles stay coupled at all times whether we use multinomial or systematic re-sampling. Maximal coupling was used in \cite{jasra2015multilevel, jasra2016multilevel} for designing efficient multi-level particle filtering algorithms and by \cite{chopin2015particle} to construct a coupling between two particle Gibbs updates from different starting points. 

Indeed, by construction, the number $C_t$ of particles that stay coupled until time $t$ decreases to zero at a much lower rate than when the independent re-sampling scheme is used; see Figure \ref{fig:proportion_paired_ricker}. The maximal coupling construction outperforms the naive independent re-sampling by orders of magnitude in most realistic scenarios and when the number of re-sampling events is small when compared to the number $N$ of particles. On the other hand, it is important to note that, even when the maximal coupling construction is used, in general and as depicted in Figure \ref{fig:proportion_paired_ricker}, the number of coupled particles does decrease exponentially fast with the number of re-sampling events.

One main drawback of the maximal coupling construction is that, as is clear from the probabilistic description \eqref{eq.max.coupl}, conditionally on the event $\{Y \neq Z\}$, the random variables $Y$ and $Z$ are independent. This means that, when used to couple two particle filter trajectories, when two particles $X^{(1)}_{i,t}$ and $X^{(2)}_{i,t}$ are decoupled at time $t$, these two particles are not close to each other in any reasonable metric. Effectively, this means that once all the particles are decoupled, which does happen after a relatively small number (typically logarithmic in $N$) of re-sampling events, there is little to no point in attempting to use the maximal coupling constructions. Indeed, once all the particles are decoupled, that is, $C_t = 0$, then for a generic index $1 \leq i \leq N$ the particles $X^{(1)}_{i,t}$ and $X^{(2)}_{i,t}$ are ``far away" from each other and attempting to drive them using the same noise is not likely to bring any efficiency gain.

It should be noted that, in some particular models, if two distinct particles are driven by the same noise process, these two particles tend to coalesce; this means that, even if two particles are decoupled, it is still worthwhile to drive them by the same noise process. In these settings, the use of the maximal coupling approach may turn out to be useful, as argued in the recent arxival \cite{jacob2016coupling}. This notion of \emph{stochastic synchronization} \cite{toral2001analytical, zhou2002noise} is extremely rare in physical systems, although it is argued in \cite{jacob2016coupling} that several widely used statistical models (e.g. autoregressive models) do enjoy this favourable contracting behaviour. Without this notion of synchronization, the gains offered by the maximal coupling approach are minor. In our numerical study, we have not observed very large gains by using the maximal coupling approach; this is because the models we have considered are not contracting.

%
%
%
%

\section{Optimal transportation re-sampling} \label{sec:OT_resampling}

In this section, we first introduce the necessary concepts related to optimal transport in Subsection \ref{sec.OT.review}. Since traditional approaches for computing optimal transport typically scale worse than quadratically in the number of particles, they are not exploitable in most modern situations where a large number of samples is required. We thus develop fast approximate transport couplings that can be used to design efficient re-sampling schemes. 
Particular care is devoted to designing algorithms that scale sub-quadratically with respect to the total number of particles.

\subsection{Optimal transport} \label{sec.OT.review}
In any non-trivial situations where one is attempting to couple two distinct particle filter trajectories, the number $C_t$ of coupled particles will decrease exponentially fast to zero; the use of the maximal coupling scheme only helps mitigate this effect. Nevertheless, one can still improve upon the maximal coupling approach. 
We would like to emphasize at this point that, through the use of the optimal transportation methodology, we are not aiming at slowing down the rate of decay of the number $C_t$ of coupled particles; indeed, numerical simulations show that the optimal transport approach to be described below leads to algorithms with a rate of decay of $C_t$ that is typically worse than what would be obtained by using the maximal coupling approach. Nevertheless, the optimal transport approach can typically generate estimates that have a much lower variance.
As explained at the end of Section \ref{sec.maximal.coupling}, the main drawback of the maximal coupling scheme is that once a pair $(X^{(1)}_t, X^{(2)}_t)$ of particles is decoupled it is not very worthwhile to try to drive $X^{(1)}_t$ and $X^{(2)}_t$ with the same noise. Indeed, this is because the maximal coupling does not take into account the locations of the particles.

In order for a coupled re-sampling scheme to be efficient, one does not necessarily need that the number $C_t$ of coupled particles remain high. What really matters is that the particles $X^{(1)}_{i,t}$ and $ X^{(2)}_{i,t}$ stay close to each other so that driving them with the same noise remains worthwhile. Consider two weighted $N$-particles systems $(X^{(1)}, W^{(1)}) \equiv \{(X^{(1)}_i,W^{(1)}_i)\}_{i=1}^N$ and $(X^{(2)}, W^{(2)}) \equiv \{(X^{(2)}_i,W^{(2)}_i)\}_{i=1}^N$. If one denotes by $\coupling(W^{(1)}, W^{(2)})$ the (convex) set of coupling matrices $\Pi$ that satisfy the constraint \eqref{eq.coupling.constraint}, the maximal coupling can also be described as a solution to the optimization problem
\begin{eqnarray*} 
\textrm{Minimize}  \bigg \{ \Pi & \mapsto & \sum_{\alpha = 1}^{N} \sum_{\beta =1}^{N} \Pi_{\alpha,\beta} \times \mathbf{1}(\alpha \neq \beta) :  \Pi \in \coupling(W^{(1)}, W^{(2)}) \bigg \}. 
\end{eqnarray*}
From this formulation, it is obvious that the locations of the particles are not taken into account. A better strategy consists in considering coupling matrices that are solutions to optimization problems that do take into account the locations of the particles; for a cost function $\cost: \mathsf{X} \times \mathsf{X} \to [0, \infty)$, we consider linear programs of the type
\begin{eqnarray*}
\textrm{Minimize} \; \bigg \{ \Pi & \mapsto & \sum_{\alpha = 1}^{N} \sum_{\beta =1}^{N} \Pi_{\alpha,\beta} \times \cost\BK{X^{(1)}_{\alpha}, X^{(2)}_{\beta}} :  \Pi \in \coupling(W^{(1)}, W^{(2)}) \bigg \}.
\end{eqnarray*}
The cost function $\cost$ penalizes the coupling of particles that are distant from each other.
This cost function is typically of the form $\cost ( X^{(1)}_{\alpha}, X^{(2)}_{\beta} ) \equiv \dist ( X^{(1)}_{\alpha}, X^{(2)}_{\beta} )^p$ for some exponent $p > 0$. This is known as optimal transport \cite{kantorovitch1958translocation} in the context of optimization theory. The optimal transport (OT) distance is 
\begin{equation} \label{eq.OT}
d_{\cost}\BK{W^{(1)}, W^{(2)}} \equiv \min\curBK{  \bra{ \Pi, \cost } \; : \; \Pi \in \coupling(W^{(1)}, W^{(2)}) }
\end{equation}
For notational convenience, and with a slight abuse of notation, we have defined the cost matrix by $\cost \in \RR^{N,N}$ by $\cost_{\alpha, \beta} = \cost ( X^{(1)}_{\alpha}, X^{(2)}_{\beta} )$ so that $\bra{ \Pi, \cost } = \sum_{\alpha = 1}^{N} \sum_{\beta=1}^{N} \Pi_{\alpha,\beta} \times \cost ( X^{(1)}_{\alpha}, X^{(2)}_{\beta} )$. 
There exist dedicated linear solvers such as the transportation simplex \cite{dantzig1998linear}, combinatorial algorithms such as the Hungarian approach \cite{kuhn1955hungarian}, and many variants thereof to solve this optimization problem. However, obtaining the exact optimal transportation solution is a computationally expensive procedure and the most efficient approaches still scale as $\OO ( N^{3} \log N )$ (e.g. \cite{pele2009fast}). 
Bringing down this cost is an active area of research and we refer to \cite{pele2009fast, cuturi2013sinkhorn, ferradans2014regularized, schmitzer2015sparse, trigila2015data} for some recent work. While the exact optimal transportation solution is expensive to obtain, it is important to note that, for the purpose of designing a coupled re-sampling scheme, it is not actually needed; it is enough to be able to efficiently build a coupling matrix $\Pi \in \coupling(W^{(1)}, W^{(2)})$ that is a reasonable  approximation to the solution of the optimal transport linear problem \eqref{eq.OT}. In Section \ref{sec.IPF}, we describe  efficient algorithms for building approximate optimal transport maps that can then be leveraged for constructing coupled re-sampling schemes.

\subsection{Iterative proportional fitting for approximate optimal transport} \label{sec.IPF} 

This section describes a strategy to efficiently build approximate optimal transport coupling matrices. 
The resulting coupling is a valid coupling in the sense that it does satisfy constraint \eqref{eq.coupling.constraint}; it is approximate since it is typically not optimal. It is worth emphasizing that the use of such an approximate optimal coupling does lead to a valid algorithm: the two coupled particle filters are marginally correctly distributed.
Our methodology is inspired by the work of \cite{cuturi2013sinkhorn}. 
The approach is based on an entropic regularization of the linear programming problem \eqref{eq.OT}.  A regularization parameter $\lambda$ is first considered and a matrix $K^{(\lambda)}_{\alpha, \beta} = \exp(-\lambda \, \cost_{\alpha,\beta})$ is constructed; the matrix exponential is to be understood entry-wise. 
The regularized problem is strictly convex and can efficiently be solved through matrix scaling algorithms to produce an approximately optimal coupling $\Pi^{\lambda}$;
in this text, we use the Sinkhorn-Knopp algorithm, a variant of the iterative proportional fitting (IPF) algorithm \cite{deming1940least}. Divisions of vectors in this section apply elementwise.

Iterative proportional fitting  is an approach that dates back to research on traffic networks in the 1930s; it was re-discovered several times in a wide variety of contexts and under several names (for example, as \\ Sheleikhovskii's method, Kruithof's algorithm, Furness method, Sinkhorn-Knopp algorithm). It is described in Algorithm \ref{algo:OT_sinkhorn}; the reader is referred to \cite{cuturi2013sinkhorn} for a more complete description. The iterative steps ensure that, at convergence, the marginals of $\Pi^\lambda$ are $W^{(1)}$ and $W^{(2)}$, while the outputted form of $\Pi^\lambda$ in terms of $K^{(\lambda)}$ ensures that large costs correspond to low probabilities. The convergence criterion in Algorithm \ref{algo:OT_sinkhorn} is chosen to be such that if $u'$ is the updated value of $u$, the iterative steps stop when $\| (u' - u)/u \| \leq 10^{-3}$.

%
%
\begin{algorithm}
\caption{Approximate Optimal Transport Using Sinkhorn Distances.}
\label{algo:OT_sinkhorn}

\begin{itemize}

\item[] \textbf{Input}: Cost matrix $\cost$, marginals $W^{(1)}$ and $W^{(2)}$, regularization parameter $\lambda > 0$. 

\item[] Construct the matrix $K^{(\lambda)}$ with $K^{(\lambda)}_{\alpha, \beta} = \exp(-\lambda \, \cost_{\alpha,\beta})$.

\item[] Initialize the positive vector $u \in \RR_+^N$ as $u = \mathbf{1}_N/N$.

\item[] \textbf{While} \textit{not converged}:

\begin{itemize}

\item[] Update $u$ by setting: 
$\qquad u \leftarrow \frac{W^{(1)}}{ K^{(\lambda)} \left [ W^{(2)} / \left \{ (K^{(\lambda)})^T u \right \}  \right ] }. $

\end{itemize}

\item[] Set $v = W^{(2)} / \{ (K^{(\lambda)})^T u \}$.

\item[] \textbf{Output}: {$\Pi^{\lambda} \equiv \diag(u) \, K^{(\lambda)} \, \diag(v)$.}

\end{itemize}
\end{algorithm}
%
%

The iterative proportional fitting algorithm is known to have a linear rate of convergence \cite{soules1991rate, knight2008sinkhorn}; if one denotes by $\Pi^{\lambda}_k$ the approximation of $\Pi^{\lambda}$ after $k \geq 1$ iterations, we have that $\left\| \Pi^{\lambda}_k - \Pi^{\lambda}\right\| \leq C \, (1-r)^k$ for some $r \in (0,1)$ and absolute constant $C >0$. Indeed, we observe in practice that one needs very few iterations of the iterative proportional fitting algorithm. This effectively means that the computation of an approximate optimal transport re-sampling scheme scales as $\OO(N^2)$. In the next section, we describe how to bring this cost further down.

\subsection{Speeding up iterative proportional fitting} \label{sec.knn}

The iterative proportional fitting algorithm requires building the matrix $K^{(\lambda)} \in \RR^{N,N}$ and carrying out matrix-vector multiplications where the matrix $K^{(\lambda)}$ is involved; the approach thus has a memory  requirement of $\OO(N^2)$ and an algorithmic complexity that scales at least as $\OO(N^2)$. To reduce this complexity, which may be undesirable  when the number of particles is large, one can exploit the fact that the non-zero entries of an optimal transport coupling matrix $\Pi \in \coupling(W^{(1)}, W^{(2)})$ typically tend to be concentrated on pairs of particles that are close neighbours; in other words, it is typically the case that $\Pi_{\alpha, \beta} = 0$ if the cost $\cost_{\alpha, \beta} = \dist(X^{(1)}_\alpha, X^{(2)}_\beta)^p$ is large. To exploit this remark, one can concentrate on finding (approximate) optimal coupling matrices $\Pi$ with non-zero entries concentrated on pairs of particles $(X^{(1)}_\alpha, X^{(2)}_\beta)$ such that $X^{(2)}_\beta$ is one of the $R$ closest neighbours of $X^{(1)}_\alpha$ among all the particles $\{X^{(2)}_i\}_{i=1}^N$, where $R \geq 1$ is a chosen threshold number of closest neighbours. For clarity, let us define the set $I_R \subset [N]^2$ as the set of pairs $(\alpha, \beta)$ of indices such that $X^{(2)}_\beta$ is one of the $R$ closest neighbours of $X^{(1)}_\alpha$ among all the particles $\{X^{(2)}_i\}_{i=1}^N$.

To find an approximate optimal transport coupling matrix $\Pi$ whose non-zero elements are in $I_R$, one can use a modified cost matrix  $\widetilde{\cost} \in \RR_+^{N,N}$ defined as
\begin{align*}
\widetilde{\cost}_{\alpha, \beta} = 
\begin{cases}
\cost_{\alpha, \beta} & \quad \textrm{if } (\alpha, \beta) \in I_R\\
+ \infty              & \quad \textrm{if } (\alpha, \beta) \not \in I_R.
\end{cases}
\end{align*}
To implement this idea in the iterative proportional fitting algorithm, by the very definition of the matrix $K^{(\lambda)} = \exp\BK{-\lambda \, \cost_{\alpha, \beta}}$, it suffices to use a modified matrix defined as $\widetilde{K}^{(\lambda)} = \exp\BK{-\lambda \, \cost_{\alpha, \beta}}$ for $(\alpha, \beta) \in I_R$ and zero everywhere else. The matrix $\widetilde{K}^{(\lambda)}$ is sparse and has only $R \times N$ non-zero elements; in practice, we observe that values $R = \OO( \log N )$ are amply enough for obtaining good approximations. The use of sparse-matrix linear algebra algorithms indeed provides huge computational savings.

To construct the sparse matrix $\widetilde{K}^{(\lambda)}$, it is necessary to compute the $R$-nearest neighbours of each of the $N$ particles. A naive approach would require a computational time that scales as $\OO(R \, N \, \log N)$. This is a well-studied problem in the computer science community; by exploiting the Euclidian structure of the state-space, methods based on KD-trees, which scale as $\OO(N \, \log N)$ \cite{wald2006building}, have been proposed. This class of methods is able to exactly compute the $R$-nearest neighbours of each particle in the cloud; this is the method that we have numerically investigated in the last part of this paper. We have used the standard KD-tree implementation available in \cite{scikit-learn}. 
It is possible to obtain further gains in efficiency by only considering approximate $R$-neighbourhoods; we refer the reader to \cite{naidan2015permutation} for approximate $R$-neighbourhood benchmarks. In high dimensions, methods based on {\it locality-sensitive hashing} \cite{slaney2008locality} usually enjoy better performances than the KD-trees approaches implemented in this article.
A KD-tree recursively partitions the particles into a multi-dimensional binary tree by cycling through the Cartesian axes, splitting each of the current set of partitions in two according to the median (in each partition) along the currently chosen axis. Finding the $R$ nearest neighbours for each $X^{(1)}_{\alpha}$ scales, once the KD-tree  is constructed, as $\OO(R \, N \, \log N)$ \cite{friedman1977algorithm}. In our experiments, for the purpose of computing $R$-neighbourhoods, we have not found the performance of the  KD-tree approach to significantly depend on the correlation structures of the cloud of particles.

%
%
%
%

\section{Numerical investigations} \label{sec:numerical_investigations}

\subsection{Ricker model} \label{sec.ricker}

We first consider a noisy non-linear ecological dynamic system. Such systems are almost invariably driven by endogenous dynamic processes plus demographic and environmental process noise and observations are corrupted by noise. Minute changes in the driving noise realization or in the system parameters can cause drastic changes in the system trajectory. In this section, we consider a simple extension of the standard Ricker model as described in \cite{wood2010statistical}. This is a $d$-dimensional non-linear state-space model; the $d$-dimensional latent process $\curBK{ X_n }_{n=0}^{T}$ is such that, for any coordinate $1 \leq i \leq d$, we have
\begin{equation} \label{eq:ricker}
X_{n+1,i} =  r \, X_{n,i} \, \exp\BK{-X_{n,i} + \epsilon_{n,i} }, ~~ X_{0,i} = x_{0,i},
\end{equation}
where the noise process $\curBK{ (\epsilon_{n,1}, \ldots, \epsilon_{n,d}) }_{n \geq 0}$ is an independent and identically distributed sequence of centred Gaussian random variables with covariance $\sigma_{\epsilon}^2 \, I_d$ and $r > 0$ is an intrinsic growth rate parameter controlling the model dynamics. The $d$-dimension observation process $\curBK{(Y_{n,1}, \ldots, Y_{n,d}) }_{n = 0} ^{T}$ is such that, conditionally upon $X_{n}$, the random variables $\{ Y_{n,1}, \ldots, Y_{n,d} \}$ are independent and $Y_{n,i} \mid X_n \sim \Poisson{\phi \, X_{n,i}}$ for a scale parameter $\phi > 0$.

In this highly non-linear setting, the maximal coupling re-sampling scheme does not perform well, even for small values of the time horizon $T$ and dimension $d$. In particular, we consider $T = 50$ observations and $d = 5$ dimensional examples. In the experiments, we have chosen a number of particles $N = 5 \times 10^3$.
 
\begin{figure*}
  \centering
\includegraphics[width=1\textwidth]{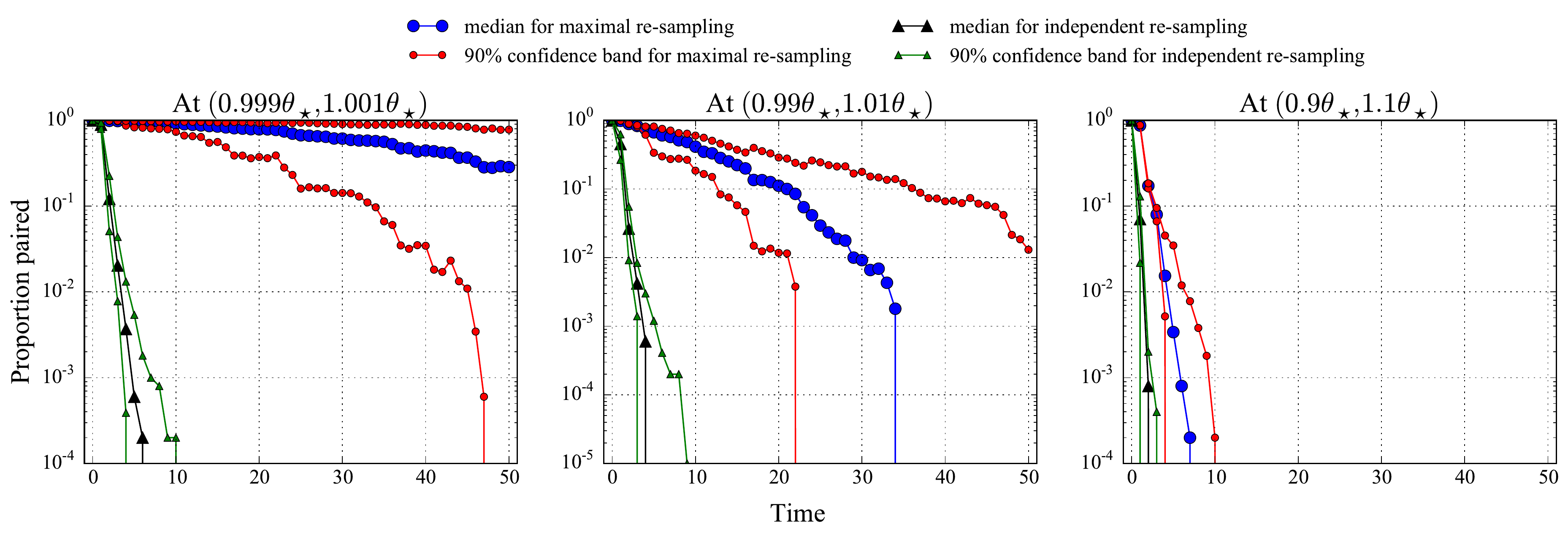}\
\caption{Proportion of particles paired for independent and maximal re-sampling in the Ricker model.}
\label{fig:proportion_paired_ricker}
\end{figure*}

For our experiments, we simulated an observation sequence $(y_0, \ldots, y_T)$ from the model itself with parameter $\theta_\star = \BK{ \log r_{\star}, \sigma_{\epsilon,\star}, \phi_{\star} } = (2, \, 0.3, \, 5)$ and initial value $x_{0,i} = 5$ for $i = 1, \ldots, d$. We coupled two particle filters, one evolving for a value of the parameter $\theta = (1-\gamma) \, \theta_\star$ and the other for a value $\theta = (1+\gamma) \, \theta_\star$; there are three simulations, one for each value of $\gamma \in \{10^{-3}, 10^{-2}, 10^{-1}\}$. Figure \ref{fig:proportion_paired_ricker} displays the proportion of paired particles $C_t / N$ for $0 \leq t \leq T$ for the independent and maximal coupling approaches. Results (not displayed) for the optimal transport, are slightly worse than the maximal coupling method in terms of rate of decay of $C_t$. As already explained, this is not especially relevant: what really matters is the ability of the methods to produce estimates with low variance, which is investigated in the next sections. In all experiments, we have performed $200$ independent runs and displayed the median as well as the $5\%$ and $95\%$ percentiles. In the simulations, re-sampling events are triggered when the effective sample size \cite{kong1994sequential} on either of the trajectories falls below $N/2$, where $N$ is the total number of particles.

As expected, the independent re-sampling scheme performs badly in all cases; as $\gamma \to 0$, the maximal coupling re-sampling scheme is able to mitigate the decrease of the proportion of paired particles. Nonetheless even for $\gamma=10^{-2}$, the number of paired particles still falls to zero after only a very small number of re-sampling events.

\begin{figure*}
  \centering
    \includegraphics[width=1\textwidth]{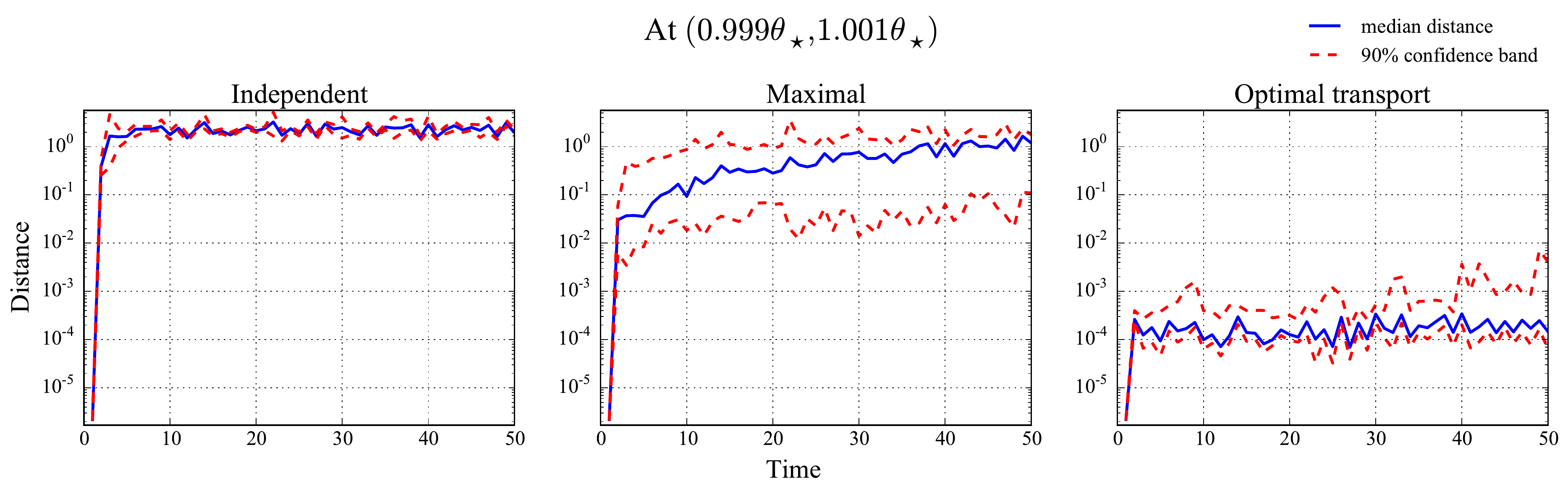}\
\caption{Average distance $E_t$ between pairs of particles for different re-sampling schemes for the Ricker model.}
\label{fig:distances_ricker_1}
\end{figure*}

In a second set of experiments, we considered the average Euclidean distances between the two populations of particles. For $0 \leq t \leq T$, we monitored the quantity $E_t \equiv  \frac{1}{N} \sum_{i=1}^{N} \|  X^{(1)}_{i,t} -  X^{(2)}_{i,t}  \|^{2}.$ For $\gamma = 10^{-3}$, which from Figure \ref{fig:proportion_paired_ricker} is the situation in which the proportion of paired particles decreases most slowly for the maximal re-sampling scheme among the considered values of $\gamma$, we compared the independent, maximal and and our proposed approximate, sub-quadratic, optimal transport based re-sampling schemes; the regularization parameter $\lambda$ for our proposed scheme was fixed at $\lambda = 50$. As displayed in Figure \ref{fig:distances_ricker_1}, even for such a small value of $\gamma$, our proposed scheme indeed outperforms the two other re-sampling schemes by orders of magnitude. It is interesting, and not too surprising in view of the discussion at the end of Section \ref{sec.maximal.coupling}, to note that the maximal coupling scheme does not yield very significant gains over the naive independent re-sampling scheme.

\begin{figure*}
\centering
\includegraphics[width=0.7\textwidth]{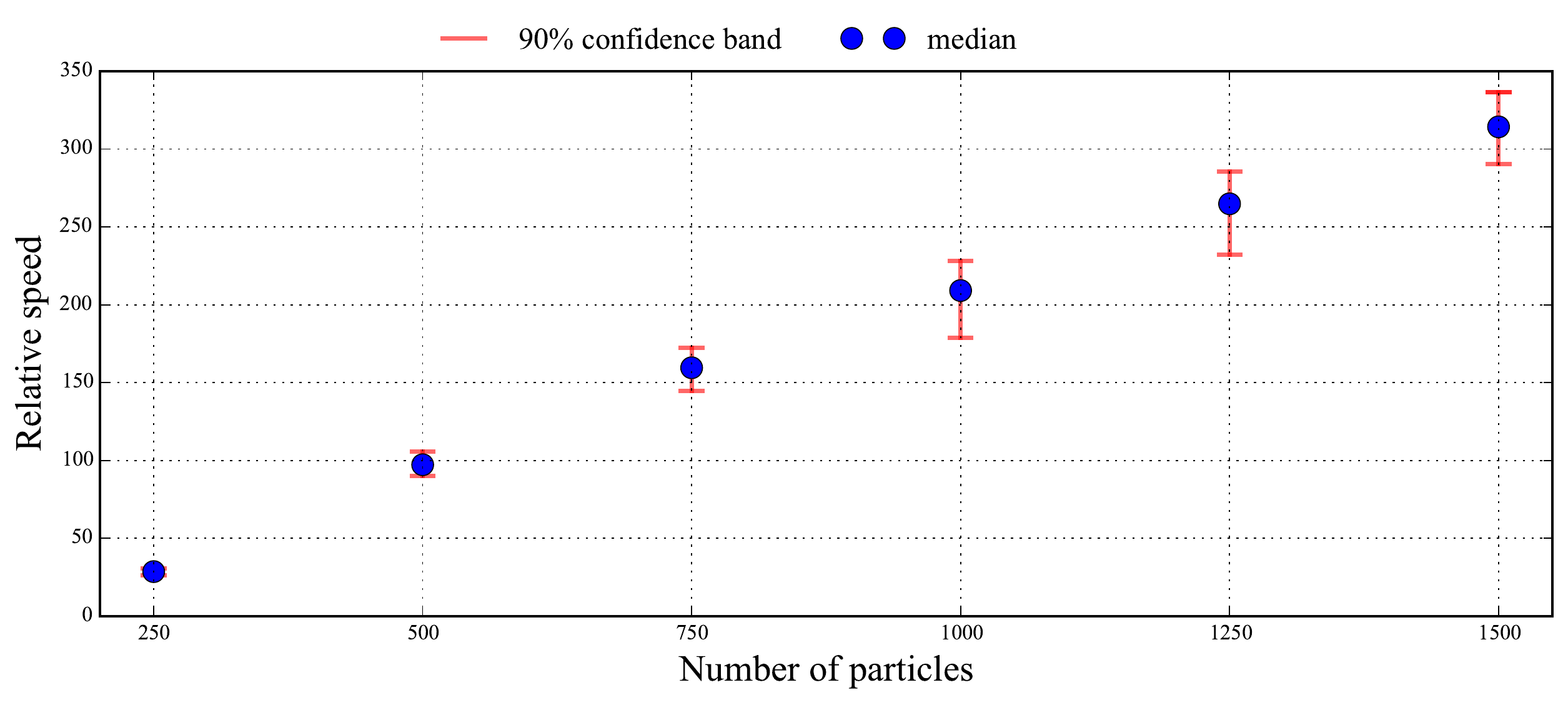}
\caption{Ratio of the time taken for usual dense matrices implementation by the time taken for the sparse matrices implementation. The median, $5\%$ and $95\%$ percentiles over $100$ independent realizations are displayed.}
\label{fig:speed_gain_sparse}
\end{figure*}

As a last set of numerical experiments, we investigated the gains in speed brought by the nearest neighbours strategy developed in Section \ref{sec.knn}. We compared the re-sampling schemes using the iterative proportional fitting relaxation to optimal transport by both dense matrices implementations (that is, Section \ref{sec.IPF}) and sparse matrices implementations (that is, Section \ref{sec.knn}). Figure \ref{fig:speed_gain_sparse} displays the gains in computational time, that is, the ratio of the time taken for usual dense matrices implementation by the time taken by the sparse matrices implementation. The nearest neighbours strategy yields orders of magnitudes of gains in computational time. The gain in speed increases as the number of particles increases; this is expected as the matrices involved become sparser as their dimension increases.

\subsection{Multi-level particle filter} \label{sec.mlpf}
We first demonstrate the benefit of using approximate optimal transport re-sampling in the context of multi-level particle filtering of \cite{jasra2015multilevel, jasra2016multilevel}. Consider the two dimensional diffusion process $\{ X_t \}_{t \geq 0}$ with multiplicative noise,
\begin{equation} \label{eq:diffusion}
dX_t = - \alpha \, X_t \, dt + \, \Gamma(\sigma X_t) \, dW_t, ~~ X_0 = x_0,
\end{equation}
where $\{W_t\}_{t \geq 0}$ is a standard Brownian motion in $\RR^2$; the volatility function $\Gamma: \RR^2 \mapsto \RR^{2,2}$ is given by
\begin{eqnarray*}
\Gamma(X) & \equiv & \begin{pmatrix} \sin R & - \cos R \\ \cos R & \sin R \end{pmatrix}
\end{eqnarray*}
with $R = \|X\|_2 = \BK{X_1^2 + X_2^2}^{1/2}$
and $\sigma > 0$ is a volatility scale parameter.
Noisy observations of the first coordinate are collected every $\delta$ unit of time and distributed as 
\begin{equation*}
Y_{k \, \delta} \mid X_{k \, \delta} \sim \Normal{X_{k \, \delta,1}, \sigma^2_{\epsilon}}
\end{equation*}
for some variance parameter $\sigma_{\epsilon} > 0$. For the numerical experiments, we consider an Euler-Maruyama discretization of equation (\ref{eq:diffusion}) 
\begin{equation} \label{eq:2d_model_Euler-Maruyama}
X_{t + \delta_t} = X_t - \alpha X_t \delta_t + \sqrt{\delta_t} \, \Gamma(\sigma \, X_t) W_t, ~~ X_0 = x_0,
\end{equation}
where $W_t \sim \Normal{0,1}$; we choose the discretization to be $\delta_t = \delta/100$. We generated observations from the model with $(\alpha_\star, \sigma_\star, \sigma_{\epsilon, \star}) = (0.5, \, 1, \, 0.5)$ and $x_0 = (0.2, 0.2)$. In this context, we use multi-level particle filtering to estimate the quantity $E[\phi(X_{k \delta}) \mid y_0, \ldots, y_{k \delta}]$ for function $\phi(x) = x_1 + x_2$, where $x = (x_1, x_2)$, and the log-likelihood $\ell(\theta \mid y_0, \ldots, y_{k \delta})$, $k = 1, \ldots, T$, for some time horizon $T$, where $\ell(\theta \mid y_0, \ldots, y_{k \delta})$ denotes the log-likelihood of the first $(k+1)$ observations.

\begin{figure*}
\centering
\includegraphics[width=\textwidth]{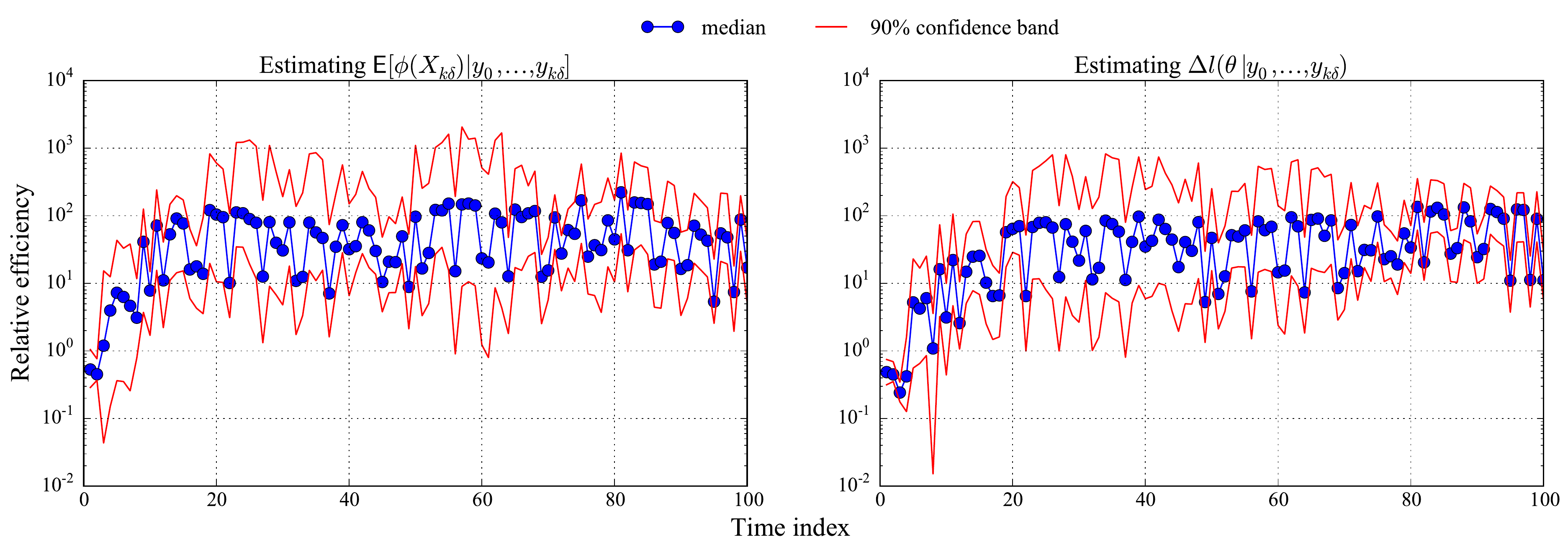}\
\caption{Multi-level particle filtering: computational inefficiency of maximal coupling divided by computational inefficiency of our proposed algorithm for estimating $E[\phi(X_{k \delta}) \mid y_0, \ldots, y_{k \delta}]$ and the delta log-likelihood for the state-space model (\ref{eq:diffusion}).}
\label{fig:filtering_mlpf}
\end{figure*} 

To keep things simple and only concentrate on the effect brought by the different re-sampling schemes (that is, avoid considering the influence of the number of particles per level), we only consider a multi-level particle filter with two levels. The first level uses a standard Euler-Maruyama discretization (\ref{eq:2d_model_Euler-Maruyama}) with discretization $\delta_t$ and the second level uses the same scheme but with discretization $\delta_t/2$. The reader is referred to \cite{jasra2015multilevel, jasra2016multilevel} for more details on multi-level particle filtering methods. In the simulations, re-sampling events are triggered when the effective sample size on either of the levels falls below $N/2$, where $N$ is the total number of particles. We have used a time horizon of $T=10^2$ with $\delta = 10^{-1}$ and parameter $(\alpha, \sigma, \sigma_{\epsilon}) = (\alpha_\star, \sigma_\star, \sigma_{\epsilon, \star})$.  The regularization parameter $\lambda$ in Algorithm \ref{algo:OT_sinkhorn} is fixed at $\lambda = 500$.

The computational inefficiency of a method is defined as the running time multiplied by the estimated variance; a lower computational inefficiency signifies better performance. We independently repeat each experiment $50$ times to get an estimate of the variance and we repeat this $25$ times independently to get a confidence interval for the estimated variance. The relative efficiency of our proposed algorithm over the maximal coupling algorithm is presented; this is defined as the ratio of the inefficiency of the maximal coupling over the inefficiency of our proposed algorithm. Figure \ref{fig:filtering_mlpf} shows that our proposed algorithm performs up to two orders of magnitude better than maximal coupling. 

\subsection{Delta log-likelihood} \label{sec:delta_loglikelihood}

\begin{figure*}
\centering
\includegraphics[width=1\textwidth]{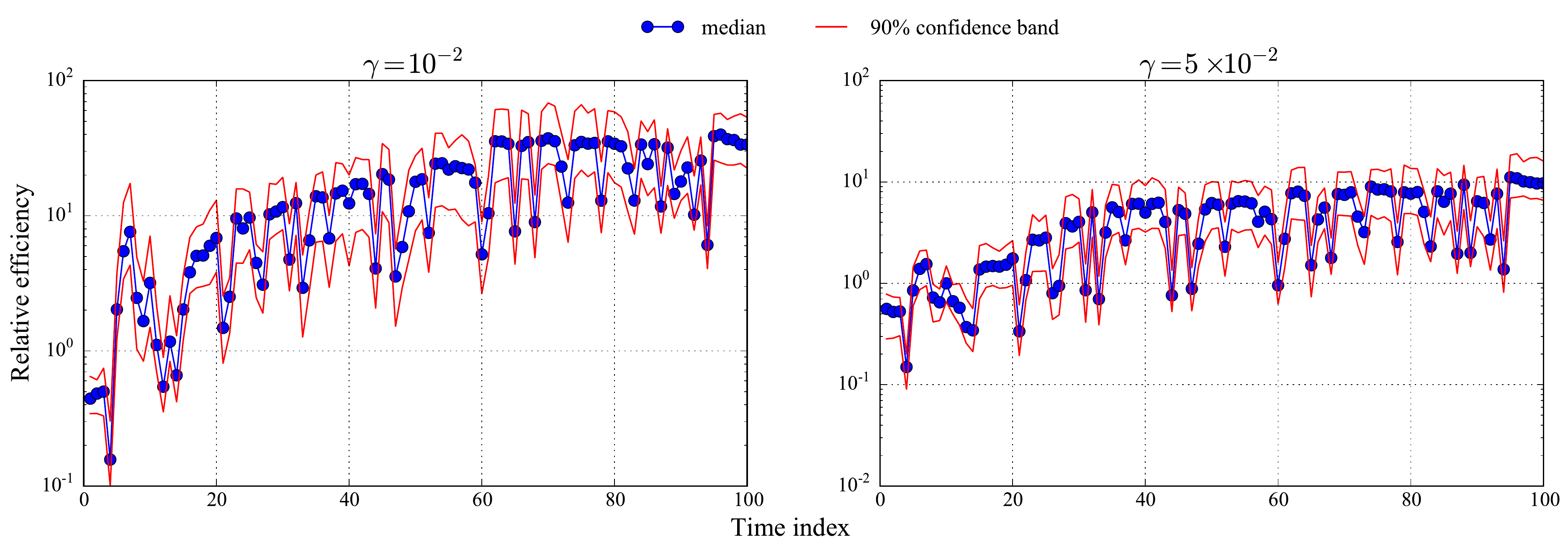}\
\caption{Delta loglikelhood: computational inefficiency of maximal coupling divided by computational inefficiency of our proposed algorithm for estimating the delta log-likelihood for the state-space model (\ref{eq:diffusion}).}
\label{fig:filtering_delta_loglikelihood}
\end{figure*}

In this section, we continue considering the state-space model \eqref{eq:diffusion} and its Euler-Maruyama discretization (\ref{eq:2d_model_Euler-Maruyama}). We denote by $\ell(\alpha, \sigma, \sigma_\epsilon) = \log \CP{ \by }{\alpha, \sigma, \sigma_\epsilon}$ the log-likelihood of the parameter $(\alpha, \sigma, \sigma_\epsilon)$; recall that the observations $\by = \BK{y_{k \, \delta}}_{k=0}^T$ are generated from \eqref{eq:2d_model_Euler-Maruyama}. We estimated the delta log-likelihood
\begin{align*}
D(\gamma) \equiv &\ell\BK{ \alpha_\star, [1+\gamma] \, \sigma_\star, \, [1+\gamma] \, \sigma_{\epsilon,\star} } 
-\ell\BK{ \alpha_\star, [1-\gamma] \, \sigma_\star, \, [1-\gamma] \, \sigma_{\epsilon,\star} }
\end{align*}
for $\gamma \in \curBK{10^{-2}, \, 5 \times 10^{-2}}$ by simply running a pair of coupled (bootstrap) particle filters for the maximal coupling and for our proposed algorithm. In the simulations, re-sampling events are triggered when the effective sample size on either of the trajectories falls below $N/2$, where $N$ is the total number of particles. Figure \ref{fig:filtering_delta_loglikelihood} shows that our proposed algorithm performs an order of magnitude better than maximal coupling.

%
%

%
%
\subsection{Markov chain Monte Carlo} \label{sec:mcmc}

In this section, we again consider the state-space model \eqref{eq:diffusion} and its Euler-Maruyama discretized version (\ref{eq:2d_model_Euler-Maruyama}).
Fixing $\sigma_{\epsilon}$ at its true value, we consider estimating parameters $\alpha$ and $\sigma$. Since the parameters $\alpha$ and $\sigma$ are positive, we consider independent random walk proposals for $\log \alpha$ and $\log \sigma$, each with standard deviation $0.5$. Log-normal priors were chosen for $\alpha$ and $\sigma$, with the prior for $\alpha$ having mean $-1$ and standard deviation $0.75$, and the prior for $\sigma$ having mean $0$ and standard deviation $0.75$. 

\begin{figure*}
\centering
\includegraphics[width=\textwidth]{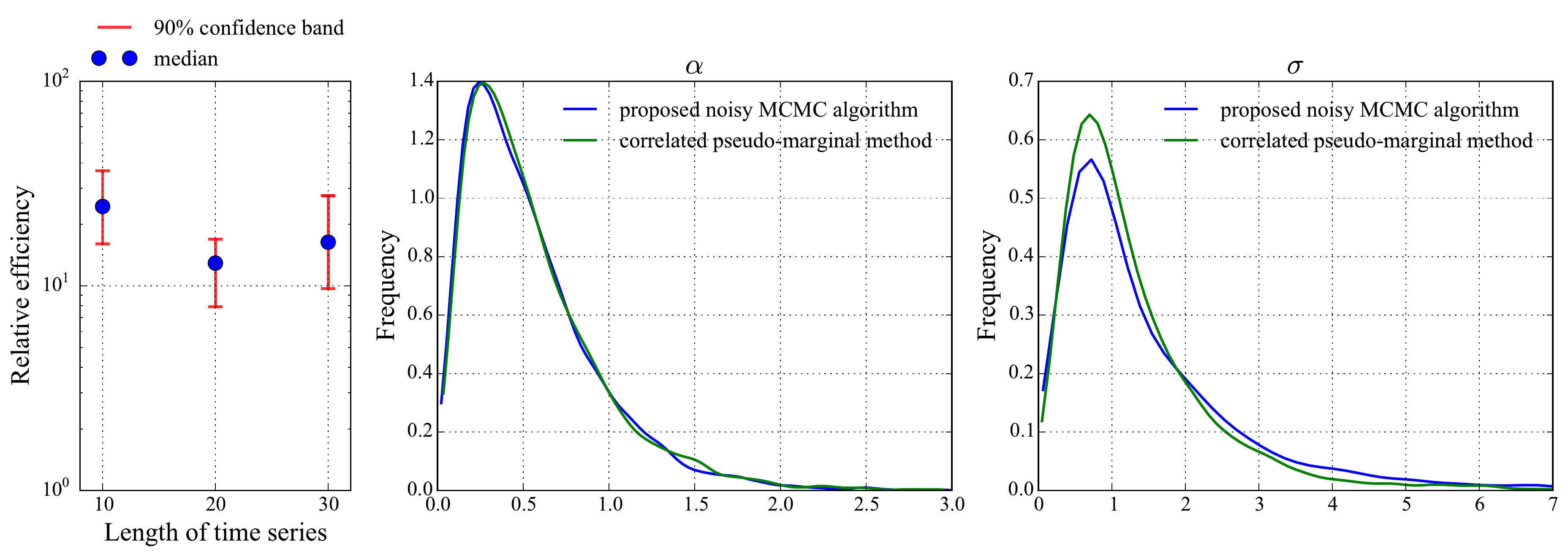} 
\caption{Comparison of our proposed noisy MCMC algorithm and the correlated pseudo-marginal method of \cite{deligiannidis2015correlated} for the state-space model (\ref{eq:diffusion}); the left panel displays the computational  inefficiency of the correlated pseudo-marginal algorithm divided by the computational  inefficiency of our proposed noisy MCMC algorithm, while the middle and the right panels display kernel density estimates for the values obtained from our proposed noisy MCMC algorithm and from the correlated pseudo-marginal method.}
\label{fig:mcmc_2d_model}
\end{figure*}

We estimate the delta log-likelihood $\Delta \, \hat{\ell}(\theta, \theta')$ in the acceptance probability (\ref{eq.mcmc_acceptance_ratio}) by running a coupled particle filter at the pair $(\theta, \theta')$ using our proposed coupled re-sampling scheme; in the simulations, re-sampling events are triggered when the effective sample size on either of the trajectories falls below $N/2$, where $N$ is the total number of particles. This leads to a so-called ``noisy'' version of a Markov chain Monte Carlo (MCMC) algorithm as the log-likelihood of the current state $\theta$ is re-estimated at each iteration and its previous estimate discarded; see \cite{franzke2015using, medina2015stability, alquier2016noisy} for some recent literature on noisy MCMC algorithms. 
It is important to emphasize that, in general, the invariant distribution of the noisy MCMC algorithm is only an approximation of the true posterior distribution of interest. Under mild assumptions, though, this discrepancy vanishes as the stochasticity of the estimates to the delta log-likelihood goes to zero. In our experiment, we have chosen a number of particles large enough such that this difference is negligible; see Figure \ref{fig:mcmc_2d_model}.

We compare our proposed noisy MCMC algorithm to the correlated pseudo-marginal algorithm of \cite{deligiannidis2015correlated}; this is an algorithm that targets the true posterior distribution of $\theta$ and correlates the auxiliary random variables used in the estimation of the log-likelihoods at each iteration of the Markov chain. The mixing parameter $\rho$ for the auxiliary random variables in the correlated pseudo-marginal algorithm was chosen to be $\rho = 0.9$, see point $2$ of the Correlated Pseudo-Marginal Algorithm (page 3) of \cite{deligiannidis2015correlated}. 
In general, even if we were using a conditional version of our coupled resampling scheme to imitate a genuine (correlated) pseudo-marginal MCMC scheme, the resulting algorithm would still not be exact. This is the main motivation for only considering noisy MCMC algorithms in this paper. We refer the reader to \cite{jacob2016coupling} for discussions and partial remedies in some particular cases.

To compare different algorithms, we consider the computational inefficiency defined as the product of the integrated auto-correlation time (IACT) with the computational time $T_{\textrm{step}}$ necessary to carry out one MCMC step
\begin{align*}
\textrm{(computational inefficiency)} \; = \;
\textrm{(IACT)} \times T_{\textrm{step}}.
\end{align*}
Comparing computational inefficiency is equivalent to comparing the MCMC variances for a fixed computational budget; this is also equivalent to the notion of effective sample size (ESS) per unit of time. To estimate the IACT, we used the method of \cite{geyer1992practical}. The MCMC algorithms were run for $10^4$ iterations and the first $10^3$ iterations were discarded. 

In order to check for the accuracy of our proposed noisy MCMC algorithm, we also plot kernel density estimates for the values obtained from the correlated pseudo-marginal algorithm and the values obtained from our proposed noisy MCMC algorithm. We present this for the case when the time series is of length $10$ and using $2 \times 10^3$ particles for the correlated pseudo-marginal method and $2 \times 10^1$ particles for our proposed noisy MCMC algorithm; similar results were obtained for longer time series. 

The results are displayed in Figure \ref{fig:mcmc_2d_model}. The left panel displays the computational inefficiency of the correlated pseudo-marginal method divided by the computational inefficiency of our proposed noisy MCMC algorithm; our proposed algorithm performs an order of magnitude better than the correlated pseudo-marginal method. The centre and right panels display the accuracy of our proposed noisy MCMC algorithm compared to the correlated pseudo-marginal method; the invariant distribution of our proposed noisy MCMC algorithm using approximate optimal transport re-sampling is very close to the correct posterior distribution obtained through a run of the correlated pseudo-marginal method.

\subsection{Prokaryotic-auto regulation} \label{sec.PAR}

Finally, in order to illustrate the performance of our method on a realistic scenario where the posterior distribution exhibits an intricate correlation structure, we consider a complex stochastic chemical kinetics model; this consists of a system of molecules of different chemical `species' reacting among themselves. In particular, we consider the prokaryotic-auto regulation model of Section $5.2$ of \cite{golightly2011bayesian}. In this model, there are four chemical species and eight reactions, with each reaction having a rate $c_i$, $i = 1, \ldots, 8$. This is approximated by a diffusion process
\begin{equation} \label{eq:PAR_model}
d X_t = \alpha(X_t,c) \, dt + \sqrt{ \beta(X_t, c) } \, d W_t, ~~ X_0 = x_0,
\end{equation}
where $\alpha(X_t, c) = S h(X_t, c)$ and  $\beta(X_t, c) = S \text{diag} \{ h(X_t,c) \} S^T$,
the four chemical species being $X = (\mathsf{DNA}, \mathsf{P_2}, \mathsf{RNA}, \mathsf{P})$.   
Here $c = (c_1, \ldots, c_8)$ are the rates of the eight reactions, $h$ is known as the hazard function which is a measure of the propensity of the reactions to occur and is given by $h(X_t,c) = (c_1 \, X_{t,1} \, X_{t,2}, \, c_2(K - X_{t,1}), \, c_3 X_{t,1}, \, c_4 X_{t,3}, \, c_5 X_{t,4}(X_{t,4}-1)/2, \, c_6 X_{t,2}, \, c_7 X_{t,3}, \, \\ c_8 X_{t,4})$, $K$ is a constant which comes from a conservation law in the model, and $S$ is known as the stoichiometry matrix and is given by 
\begin{align*}
S = 
\begin{pmatrix} 
0 & 0 & 1 & 0 & 0 & 0 & -1 & 0 \\ 
0 & 0 & 0 & 1 & -2 & 2 & 0 & -1 \\ 
-1 & 1 & 0 & 0 & 1 & -1 & 0 & 0 \\ 
-1 & 1 & 0 & 0 & 0 & 0 & 0 & 0 
\end{pmatrix}. 
\end{align*}
Observations are available at every unit of time and are distributed as $Y_t = (0, 1, 2, 0) X_t + \epsilon_t$, where $\epsilon_t \sim \Normal{0, \sigma_\epsilon^2}$, $t = 1, \ldots, T$. We consider an Euler-Maruyama discretization of the process (\ref{eq:PAR_model}), 
\[ X_{t + \delta_t} = X_t + \alpha(X_t,c) \, \delta_t + W_t, ~~ X_0 = x_0, \]
where $W_t \sim \Normal{ \mathbf{0}_4, \delta_t \,\beta(X_t,c) }$ is a four-dimensional Gaussian random variable with mean $\mathbf{0}_4$ and and covariance matrix $\delta_t \, \beta(X_t,c)$. We refer the reader to \cite{gillespie2007stochastic, erdi2014stochastic} for a more complete description of stochastic chemical kinetic models.

\begin{figure*}
\centering
\includegraphics[width=1\textwidth]{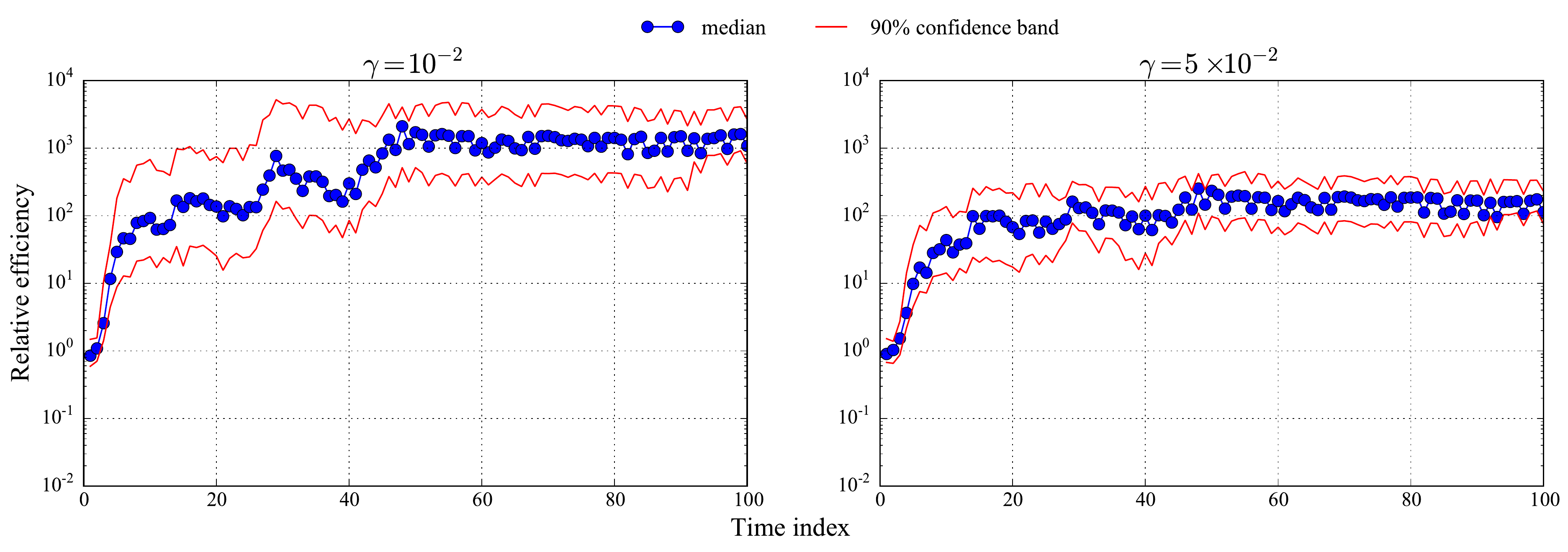} 
\caption{Computational inefficiency of maximal coupling divided by computational  inefficiency of our proposed algorithm for estimating the delta log-likelihood for the prokaryotic auto-regulation model (\ref{eq:PAR_model}).}
\label{fig:PAR_delta_loglikelihood}
\end{figure*}

The initial value of the process is chosen to be $x_0 = (8, 8, 8, 5)$, the true reaction rates to be $c_{\star} =$ (0.1, 0.7, 0.35, 0.35 0.2, 0.1, 0.9, 0.3, 0.1), the Euler-Maruyama discretization to be $\delta_t = 1/10$, the time horizon to be $T = 100$ and the variance of the noise term to be $\sigma_\epsilon^2 = 10$. For our experiments, we fix $c_3, \ldots, c_8$ at their true values and consider the delta log-likelihood
\begin{align*}
D(\gamma) &= \ell( [1+\gamma] c_{1,\star}, [1+\gamma] c_{2,\star}, c_{3,\star}, \ldots, c_{8,\star}) - \ell( [1-\gamma] c_{1,\star}, [1-\gamma] c_{2,\star}, c_{3,\star}, \ldots, c_{8,\star})
\end{align*}
for $\gamma \in (10^{-2}, 5 \times 10^{-2})$. In the simulations, re-sampling events are triggered when the effective sample size on either of the trajectories falls below $N/2$, where $N$ is the total number of particles. We again compare the relative efficiencies of the maximal coupling and our proposed algorithm with regularization parameter $\lambda$ fixed at $\lambda = 50$. Figure \ref{fig:PAR_delta_loglikelihood} shows that our proposed algorithm performs up to three orders of magnitude better than maximal coupling.  

\section{Conclusion} \label{sec:conclusion}

The standard approach to couple two particle filters is to drive the particles with the same noise process; in practice, this is done by setting the so-called ``random seed" to the same value before running each one of the particle filters. We demonstrate in this text that a careful coupled re-sampling step is crucial, when coupling two particle filter trajectories, to obtain good performances. While algorithms based on the maximal coupling approach can sometimes enjoy improved performances, we argue in this text that more advanced coupling methods based on ideas extracted from the optimal transportation literature can yield algorithms that are orders of magnitude more efficient. Importantly, we have described how to reduce the cost of standard optimal transportation algorithms by leveraging fast algorithms for finding nearest neighbours in populations of particles. 
As a final note, we mention that, while it was not something we observed in our experiments, if the two set of particles have vastly different ranges, then directly using a KD-tree to find nearest neighbours may not work well. In this case, the creation of an alternative metric and using it to pre-process the particles could be beneficial. For instance, the Mahalanobis distance \cite{mahalanobis1936generalized} could be used, which can be obtained in $\mathcal{O}(N d^3)$, where $N$ is the number of particles used and $d$ is the dimension of the state-space. Since we do not consider high-dimensional examples, this will not prove to be a computational bottleneck.\\

\noindent
{\bf Acknowledgements:} we would like to thank two anonymous reviewers for their constructive comments. Ajay Jasra was funded by AcRF tier 1 grant R-155-000-156-112. Alexandre Thiery was funded by AcRF grant R-155-000-150-133.

\bibliographystyle{alpha}      
\bibliography{references}   

\end{document}